\documentclass[twocolumn,showpacs,superscriptaddress,amsmath,amssymb,prl]{revtex4}

\usepackage{graphicx}%
\usepackage{dcolumn}%

\begin{document}

\title{Manifestation of the magnetic resonance mode in the nodal quasiparticle lifetime in superconducting cuprates}

\author{A. A. Kordyuk}
\affiliation{Leibniz-Institut f\"ur Festk\"orper- und Werkstoffforschung Dresden, P.O.Box 270016, 01171 Dresden, Germany}
\affiliation{Institute of Metal Physics of National Academy of Sciences of Ukraine, 03142 Kyiv, Ukraine}

\author{S. V. Borisenko}
\affiliation{Leibniz-Institut f\"ur Festk\"orper- und Werkstoffforschung Dresden, P.O.Box 270016, 01171 Dresden, Germany}

\author{A. Koitzsch}
\author{J. Fink}
\author{M. Knupfer}
\author{B. B\"uchner}
\affiliation{Leibniz-Institut f\"ur Festk\"orper- und Werkstoffforschung Dresden, P.O.Box 270016, 01171 Dresden, Germany}

\author{H. Berger}
\author{G. Margaritondo}
\affiliation{Institut de Physique de la Mat\'erie Complex, Ecole Politechnique F\'ederale de Lausanne, CH-1015 Lausanne, Switzerland}

\author{C. T. Lin}
\author{B. Keimer}
\affiliation{Max-Planck Institut f\"ur Festk\"orperforschung, D-70569 Stuttgart, Germany}

\author{S. Ono}
\author{Yoichi Ando}
\affiliation{Central Research Institute of Electric Power Industry, Komae, Tokyo 201-8511, Japan}

\date{January 1, 2004}%

\begin{abstract}
Studying the nodal quasiparticles in superconducting cuprates by photoemission with highly improved momentum resolution, we show that a new "kink" feature in the scattering rate is a key to uncover the nature of electron correlations in these compounds. Our data provide evidence that the main doping independent contribution to the scattering can be well understood in terms of the conventional Fermi liquid model, while the additional doping dependent contribution has a magnetic origin. This sheds doubt on applicability of a phonon-mediated pairing mechanism to high temperature superconductors. 
\end{abstract}


\preprint{\textit{xxx}}

\maketitle

The concept of self-energy, viz.~a non-local dynamic potential that encapsulates the effects of the electronic correlations on the behaviour of an individual electron, plays a fundamental role in many-body physics \cite{AGD}. Direct access to this quantity, offered nowadays by angle-resolved photoemission spectroscopy (ARPES) \cite{DamascelliRMP03}, has stimulated attempts to answer questions that are vital for the understanding of high-temperature superconductivity: (i) "what is responsible for the unusual normal state properties?", and (ii) "what couples electrons in pairs in the superconducting state?". The self-energy of nodal quasiparticles determined by ARPES \cite{VallaSci99, BogdanovPRL00, KaminskiPRL01, JohnsonPRL01, LanzaraNature01} has turned out to be a key quantity to examine both problems which can be reduced to the validity of the marginal Fermi liquid phenomenology (MFL) \cite{VallaSci99, VarmaPRL89, LittlewoodPRB92, HaslingerEPL02} and to the origin of the "kink" in the dispersion \cite{BogdanovPRL00, KaminskiPRL01, JohnsonPRL01, LanzaraNature01}, respectively. 

The photoemission intensity, which is measured by ARPES as a function of the kinetic energy and inplane momentum of the outgoing electrons, provides access to the spectral function of a single electron removal which is supposed to reflect the quasipartical properties of the remaining photohole: its effective mass and life time. These properties can be expressed in terms of a quasiparticle self-energy \cite{AGD} $\Sigma(\omega) = \Sigma'(\omega) + i \Sigma''(\omega)$. The MFL phenomenology has been introduced in order to describe many unusual physical properties of the normal state in superconducting cuprates \cite{VarmaPRL89}.  One of the main present arguments in favour of the "marginality" of quasiparticles in cuprates came from photoemission spectra taken along the nodal direction of the Brillouin zone (BZ), namely from analysis of the scattering rate which is proportional to the imaginary part of the self-energy. It has been shown that $\Sigma''(\omega,T)$ within a certain frequency range can be well approximated by a linear dependence on frequency and temperature \cite{VallaSci99} that agrees with the MFL model, although it has been pointed out that such an agreement could be accidental \cite{HaslingerEPL02}. The peculiarities in the photoemission spectra which appear upon entering the superconducting state are commonly ascribed to the interaction with the magnetic mode (see \onlinecite{FongPRB00, EschrigPRB03} and references therein), but recently a phonon mechanism has been revived \cite{LanzaraNature01, ZhouNature03}. Again, the focusing point is the nodal direction, where the quasiparticle dispersion exhibits a so-called "kink" \cite{BogdanovPRL00, KaminskiPRL01}. The fact that the kink in the experimentally observed dispersion hardly depends on doping and persists in all families of superconducting cuprates \cite{LanzaraNature01, ZhouNature03} could be a solid argument for the phonon scenario. However, one can argue that the dispersion kink is not the best quantity to monitor small changes in the electron self-energy because it appears as just a sharpening of a bend of the same sign in the renormalized (experimental) dispersion which is present at any temperature and doping \cite{JohnsonPRL01, KoitzschXXX03}. The kink in the scattering rate, which has been recently reported for La$_{2-x}$Sr$_x$CuO$_{4+\delta}$  \cite{ZhouNature03} and Bi$_{2-x}$Pb$_x$Sr$_2$CaCu$_2$O$_{8+\delta}$  \cite{KoitzschXXX03}, appears to be much more convenient in this sense because it develops on top of the strong normal state scattering which has the opposite curvature. 

In this Letter we show that the "scattering rate kink" makes it possible to distinguish between the different scattering channels. We argue that the main contribution to the scattering can be well understood in terms of the conventional Fermi liquid model (FL) \cite{AGD} while the additional doping dependent contribution apparently has a magnetic origin \cite{FongPRB00, EschrigPRB03}.

It is the width of the momentum distribution curves (MDCs) which is unambiguously related to the scattering rate: $\Sigma''(\omega) = v_b(\omega)$ FWHM$(\omega) / 2$, where $v_b$, is the bare velocity whose frequency dependence we neglect here, taking $v_b(\omega) \approx v_F$ = 4 eV\AA~\cite{KordyukPRB2003}, and FWHM$(\omega)$ is the full width at half maximum of each MDC at given $\omega$. Considerable improvement of the momentum resolution, which allows to observe a FWHM for $E_F$ MDC of 0.015 \AA$^{-1}$, gave us the possibility to perform a study on the scattering rate kink as a function of doping and temperature. We investigated the superstructure-free Bi$_{2-x}$Pb$_x$Sr$_2$CaCu$_2$O$_{8+\delta}$ [Bi(Pb)-2212] in a wide doping range, as well as Bi$_2$Sr$_2$CaCu$_2$O$_{8+\delta}$ (Bi-2212) and Bi$_2$Sr$_{2-x}$La$_x$CuO$_{6+\delta}$ (Bi-2201). We label the samples according to their doping level and critical temperature, e.g., "UD76" reads for underdoped, $T_c$ = 76 K. The doping level of the samples has been checked using the Fermi surface mapping technique \cite{KordyukPRB2002} at room temperature and measuring the coupling strength \cite{KimPRL2003} below $T_c$. The presented data have been obtained using an experimental setup where we combined a high resolution light source with a wide excitation energy range (U125/1-PGM beamline at BESSY), an angle-multiplexing photoemission spectrometer and a 3-axis rotation cryo manipulator. Bi(Pb)-2212 OP88 has been measured using a He discharge lamp.

\begin{figure}[t]
\includegraphics[width=6.5cm]{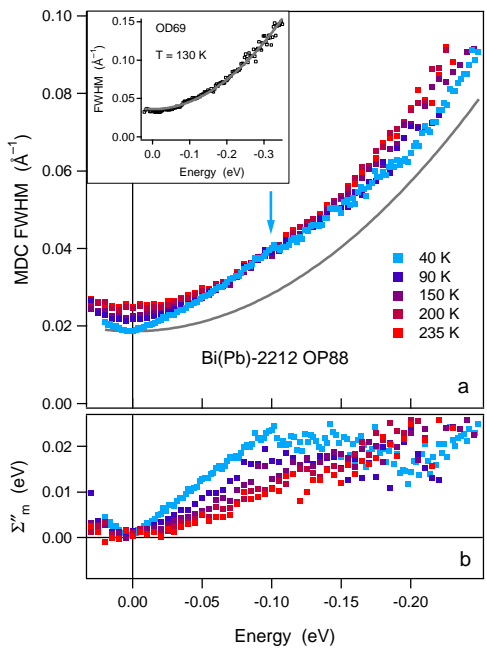}%
\caption{\label{bud}  
Temperature dependence of the scattering rate for the nodal quasiparticles in optimally doped (Bi,Pb)$_2$Sr$_2$CaCu$_2$O$_{8-\delta}$. In a, the full width at half maximum (FWHM) of the momentum distribution curves (MDCs) of the photoemission intensity is shown for different temperatures as a function of the energy of the quasiparticles. The gray solid line represents a contribution from the usual Auger decay (Fermi liquid parabola) \cite{AGD} obtained by fitting the data for highly overdoped sample (OD69) at 130 K (see inset). b illustrates the result of a subtraction of the Fermi liquid parabola for each temperature in terms of the imaginary part of the self-energy (the FWHM/2 is multiplied to the bare Fermi velocity $v_F$ = 4 eV\AA).}
\end{figure}

Figure 1a shows the scattering rate as a function of frequency for optimally doped Bi(Pb)-2212 OP89 for different temperatures. The scattering rate is presented in momentum units. A sharp kink seen in $\Sigma''(\omega)$ at 0.1 eV (indicated by the arrow) at 40 K (below $T_c$ = 88 K) gradually vanishes with increasing temperature. Another important finding is that the high binding energy tail of $\Sigma''(\omega)$ shifts upwards with temperature similar to the $\Sigma''(0)$ value. This shift, being in agreement with optical conductivity results \cite{vanderMarelNature03}, contradicts, in fact, the MFL scenario \cite{VarmaPRL89}, according to which $\Sigma''(\omega, T) \propto \max(|\omega|, T)$. Such a shift of the whole curve is expected within the FL model when the scattering rate is determined by an Auger-like decay (the process where, in our case, the hole decays into two holes and one electron \cite{AGD}) that gives $\Sigma'' \propto \omega^2 + (\pi T)^2$. The FL behaviour is generally expected for overdoped samples \cite{VarmaPR02}, and in Fig. 1a we add the FL parabola (solid line) which perfectly fits the scattering rate for an OD69 sample above $T_c$ in the whole binding energy range. This parabola evidently describes the main contribution to $\Sigma''$ at any temperature. The additional contribution, which is seen as a hump on top of the FL parabola, must originate from an additional interaction which can be responsible for the unusual properties of the cuprates. In Figure 1b we evaluate this interaction subtracting the FL parabola for each temperature and setting the resulting offsets to zero. We ascribe this additional interaction to scattering via a bosonic channel for reasons which we discuss below. 

\begin{figure*}[tb]
\includegraphics[width=15cm]{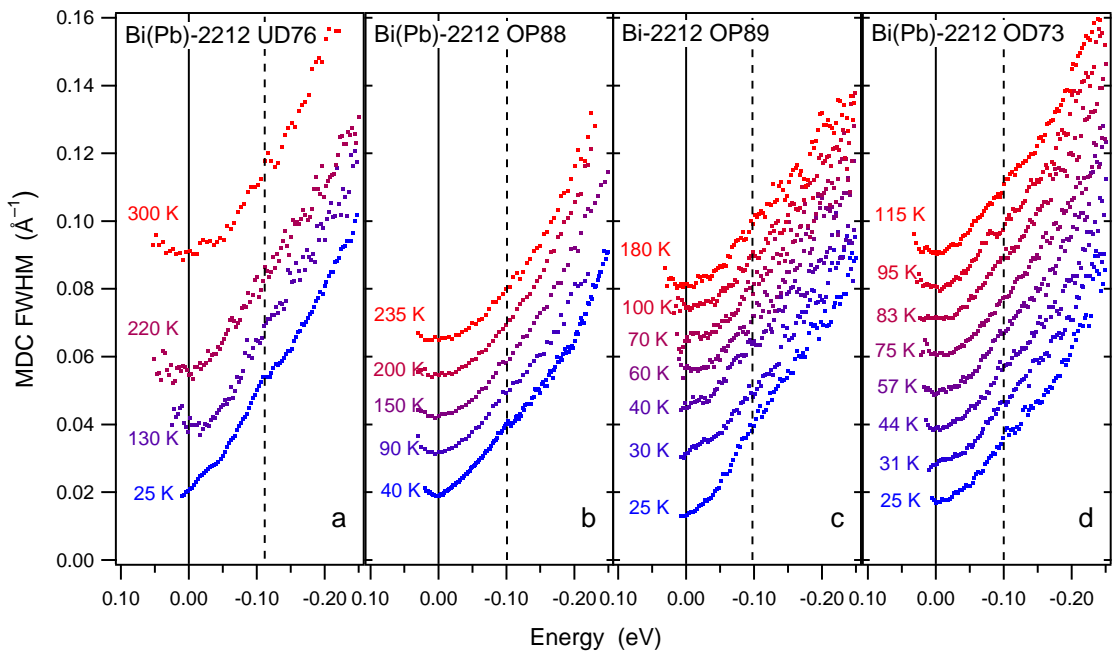}%
\caption{\label{Water} Evolution of the scattering rate kink with doping and temperature. The MDC width as function of energy is presented for a selected number of temperatures for the superstructure free Bi(Pb)-2212 and pure Bi-2212 samples of different doping levels from underdoped, $T_c$ = 76 K (a), to overdoped, $T_c$ = 73 K (d).}
\end{figure*}

Two components are essential for the scattering process via a bosonic channel: the boson density of states, which supplies bosons to scatter with, and the electron density of states, which supplies a phase space for electrons to scatter into. Any peculiarities like peaks or gaps in both densities can result (alone or in their combination) in the discussed kink on $\Sigma''(\omega)$. For example, the superconducting gap or pseudogap can produce a drop in the scattering rate at low binding energy $(|\omega| < \Delta)$ even within the Auger scattering process. A similar kink can be expected from a van Hove singularity which is close to the Fermi level in overdoped cuprates \cite{KordyukPRL2002, BorisenkoPRL2003}. On the other hand, a bosonic mode (either phononic or magnetic) or a bosonic gapped spectrum, even with a constant electronic density of states, can be responsible for such an additional scattering channel \cite{EschrigPRB03} displayed in Fig. 1b--in the simplest case of a constant electronic density of states the coupling to a single mode gives a step function in $\Sigma''(\omega)$. In order to distinguish between these possible reasons for the scattering rate kink formation we studied the $\Sigma''(\omega)$ dependencies as function of doping and temperature.

Figure 2 shows a set of  $\Sigma''(\omega)$ curves (waterfall plots with 0.01 \AA$^{-1}$ offset) for Bi(Pb)-2212 UD76 (a), OP88 (b), OD73 (d) and Bi-2212 OP89 (c) at different temperatures. While for the overdoped sample the kink disappears above $T_c$ (between 57 K and 75 K in Fig. 2d), for the underdoped samples it persists above $T_c$ and vanishes at higher temperatures which may be related to $T^*$. In Figure 3 we compare the absolute values of $\Sigma''(\omega)$ for underdoped (UD76) and overdoped (OD73) Bi(Pb)-2212 at $T$ = 25 K. The room temperature scattering rates for these two samples coincide within the experimental error bars.  It is seen that at low temperature the underdoped sample exhibits a much higher scattering rate with a more pronounced kink that has a tendency to disappear completely at higher doping levels \cite{KoitzschXXX03}. The differences between these data and the FL parabola (solid line, the same as in Fig. 1a) demonstrate that the additional scattering channel of the nodal quasiparticles is highly doping dependent which rules out the phonon scenario, leaving space for magnetic excitations as the only bosons responsible for this additional channel \cite{FongPRB00, EschrigPRB03}. 

\begin{figure}[!b]
\includegraphics[width=6cm]{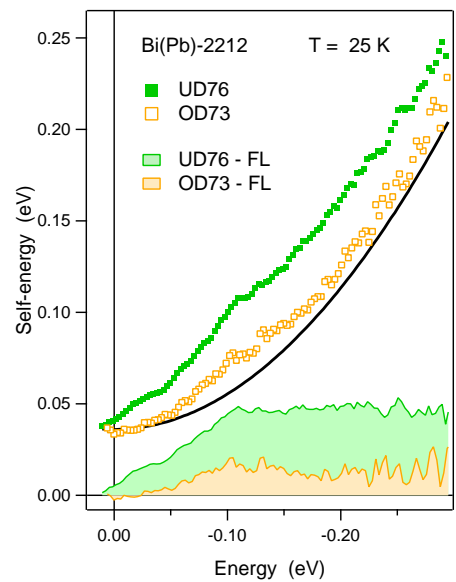}%
\caption{\label{Doping} Strengthening of the scattering mode with underdoping. Comparison of the imaginary part of the self-energy of nodal quasiparticles in Bi(Pb)-2212 underdoped ($T_c$ = 76 K) and overdoped ($T_c$ = 73 K) samples at 25 K. The black solid line represents a contribution from the Fermi liquid parabola. The shaded areas represent the contributions from the magnetic scattering obtained by subtraction of the FL parabola.}
\end{figure}

According to the magnetic scenario two contributions can be distinguished: scattering with the spin-fluctuation mode and with the spin-fluctuation continuum. Since the electron density of states should peak at the gap energy, $\Delta$, in the superconducting state and at the van Hove singularity, $E_M$, in both superconducting and normal states, the scattering associated with the mode $\Omega_{res}$ is expected to peak in between $\Omega_{res}+\Delta$ and $\Omega_{res}+E_M$, while the gapped continuum is expected to contribute at energies above 0.1 eV \cite{EschrigPRB03}. The presented results can be understood as follows: the contribution of the mode gradually increases with lowering temperature (Fig. 1b) while both contributions from the mode and continuum increase with underdoping (Fig. 3). The essential contribution from the spin-fluctuations in underdoped samples may result in a linear dependence of $\Sigma''(\omega)$, which had been considered as a manifestation of the MFL scenario.

The evidence that it is an additional channel, and not just a modification of the electronic density of states due to pseudogap opening, comes not only from comparison of the low temperature $\Sigma''(\omega)$ curves in Fig. 3 but also from the absolute value of the energy scale of the observed kink -- at about 0.1 eV and only slightly dependent on doping (see Fig. 2), i.e. it cannot be explained by a gapped electronic density of states alone but only in combination with the spin-fluctuation mode and/or gapped spin-fluctuation continuum: for the values $\Omega_{res}+\Delta$ and $\Omega_{res}+E_M$ the increase of $\Delta$ and $\Omega_{res}$ with underdoping can be compensated by the decrease of $\Omega_{res}$.
 
\begin{figure}[!t]
\includegraphics[width=8.5cm]{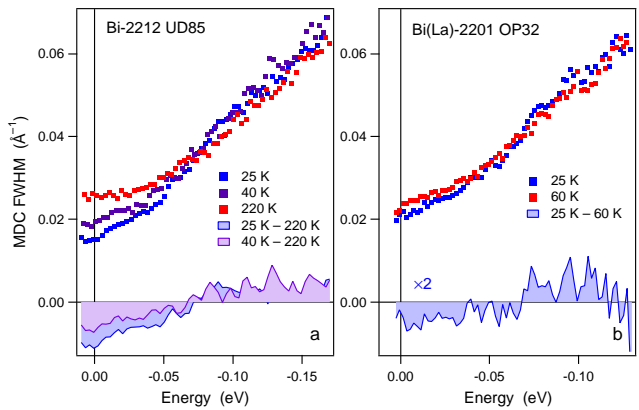}%
\caption{\label{Bi2201} Sharpening of the scattering rate kink with lowering temperature for double- and single-layer compounds.  The panels show the data for temperatures below and above the superconducting transitions as well as their difference for underdoped ($T_c$ = 85 K) Bi-2212 (a) and optimally doped ($T_c$ = 32 K) Bi(La)-2201 (b) (for Bi(La)-2201 the difference is multiplied by factor of 2).}
\end{figure}

An example where with lowering temperature the increase of the magnetic contribution overcomes the decrease of the contribution from the Auger decay in the energy range 0.07 eV $< |\omega| <$ 0.15 eV is presented in Fig. 4a for the Pb-free UD85. We explain it by the rapid sharpening of the mode below $T_c$. A similar but rather weak effect we also observe for a single-layer Bi-2201 for which the observation of the neutron resonance mode has not been reported yet.  Figure 4b shows the appearance of the scattering rate kink below the superconducting transition temperature ($T_c$ = 32 K). The kink here has a similar energy scale (about 90 meV) that allows us to conclude that a spin fluctuation mode is present in Bi-2201, but that its spectral weight is significantly smaller than in Bi-2212.

In conclusion, we have shown that the "scattering rate kink", a new feature seen in the frequency dependence of the imaginary part of the self-energy by ARPES, makes it possible to distinguish between the different scattering channels. The main contribution to the scattering can be well understood in terms of the conventional Fermi liquid model while the additional doping dependent contribution apparently has a magnetic origin. The latter manifests itself in the doping and temperature dependence. On the top of the usual Auger decay, even for the nodal quasiparticles, the magnetic contribution essentially increases with underdoping becoming dominant for the rest of the Brillouin zone and therefore determines the unusual properties of the cuprates in the superconducting and pseudo-gap phases.

We acknowledge the stimulating discussions with A. Chubukov, I. Eremin and R. Hayn, technical support by R. Follath, R. H\"ubel, S. Leger and M. R\"ummeli. H.B. and G.M. are grateful to the Swiss National Science Foundation and its NCCR Network "Materials with Novel Electronic Properties" for support.

\end{document}